\documentclass[twocolumn, secnumarabic, amssymb, amsmath,nobibnotes, aps, prl,nofootinbib]{revtex4}

\usepackage{graphicx}
\usepackage{tabularx}

\newcommand{\twothirds}{\mbox{\small{$\frac{2}{3}$}}}
\newcommand{\third}{\mbox{\small{$\frac{1}{3}$}}}
\newcommand{\fourthirds}{\mbox{\small{$\frac{4\pi}{3}$}}}
\newcommand{\fivethirds}{\mbox{\small{$\frac{5}{3}$}}}
\begin{document}
\title{Errors in the Bag Model of Strings, \\ and Regge Trajectories Represent the Conservation of Angular Momentum in Hyperbolic Space}
\author{B. H. Lavenda}
\email{info@bernardhlavenda.com}
\homepage{www.bernardhlavenda.com}
\affiliation{Universit$\grave{a}$ degli Studi, Camerino 62032 (MC) Italy}
\begin{abstract}
The MIT bag model is shown to be wrong because the bag pressure cannot be held constant, and the volume can be fixed in terms of it. The bag derivation of Regge's trajectories is invalidated by an integration of the energy and angular momentum over  all values of the radius up to $r_0=c/\omega$. This gives the absurd result that \lq total\rq\ angular momentum decreases as the frequency increases. The correct expression for the angular momentum is obtained from hyperbolic geometry of constant negative curvature $r_0$. When the square of the relativistic mass is introduced, it gives a negative intercept which is the Euclidean value of the angular momentum. Regge trajectories are simply statements of the conservation of angular momentum in hyperbolic space. The frequencies and values of the angular momentum are in remarkable agreement with experiment.
\end{abstract}
\maketitle
\section{Bag Model and Quark Confinement}
The bag model allows the mass of hadrons to be determined in terms of the masses of quarks. That is, the model relates quark confinement to the low-lying states of hadrons. The model assumes that quarks are confined to a bubble of radius $r$ of the \lq empty\rq\ vacuum upon which a constant pressure, $p$, of the \lq true\rq\ vacuum is exerted~\cite{AC}. 

The internal energy is given by
\begin{equation}
E=\frac{\hbar^2}{2mr^2}, \label{eq:E-non}
\end{equation}
where $m$ is the mass of the quark. If the compressional work, $\fourthirds pR^3$, and the masses of the three quarks are added onto \eqref{eq:E-non}, we come out with the mass of the proton~\cite{GW},
\begin{equation}
m_p=3m+\frac{\hbar^2}{2mr^2}+\fourthirds pr^3. \label{eq:M-non}
\end{equation}
The condition that the mass of the proton  be stationary with respect to the radius of the bag fixes the latter at
\begin{equation}
r_{\min}=\left(\frac{\hbar^2}{4\pi mp}\right)^{1/5}. \label{eq:Rmin}
\end{equation}
Hence, the conclusion is reached that the internal energy,
\[
E=\fivethirds\frac{\hbar^2}{2mr_{\min}^2}, \]
is $\fivethirds$ times the kinetic energy, \eqref{eq:E-non}. One may rightly question where did all this come from.

Consider the doctrine of latent, $L_V$, and specific, (here written as the heat capacity, $C_V$) heats
\begin{equation}
dQ=L_VdV+C_VdT, \label{eq:dQ}
\end{equation}
to which we subtract the work of compressional forces to obtain the increment in the internal energy,
\begin{equation}
dE=dQ-dW=\left(L_V-p\right)dV+C_VdT. \label{eq:dE}
\end{equation}
Employing the Clapeyron equation,
\begin{equation}
\left(\frac{\partial p}{\partial T}\right)_V=\frac{L_V}{T}, \label{eq:Clap}
\end{equation}
we can write the partial derivative of $E$ with respect to $V$ as
\begin{equation}
\left(\frac{\partial E}{\partial V}\right)_T=T\left(\frac{\partial p}{\partial T}\right)_V-p. \label{eq:LV}
\end{equation}
Evaluating this equation with the aid of the equation of state,
\begin{equation}
sE=pV, \label{eq:Grun}
\end{equation}
which follows directly from the virial, where $s\in[\third,\twothirds]$ is the Gr\"uneisen parameter in $3$ dimensions, we obtain the partial differential equation
\begin{equation}
E=T\left(\frac{\partial E}{\partial T}\right)_V-\frac{V}{s}\left(\frac{\partial E}{\partial V}\right)_T. \label{eq:de}
\end{equation}

The partial differential equation \eqref{eq:de} can be solved by the method of characteristics by integrating the simpler, auxiliary equations
\[\frac{dT}{T}=-s\frac{dV}{V}=\frac{dE}{E}.\]
This leads to two independent solutions $a=TV^s$, and $b=EV^s$, which can be expressed as
\begin{equation}
\Phi(a,b)=0 \qquad \mbox{or}\qquad E=V^{-s}\phi\left(TV^s\right),\label{eq:phi}
\end{equation}
where $\Phi$ and $\phi$ are completely arbitrary functions. Now there are two possibilities as $T\rightarrow0$~\cite{HE}:
\begin{enumerate}
\item from the second independent solution 
\begin{equation}
E=CV^{-s},\qquad \left(\frac{\partial E}{\partial V}\right)_T<0,\label{eq:repulsive}
\end{equation}
 so the system will have a vanishing latent heat in \eqref{eq:LV}, or
\item from the equation of state, \eqref{eq:Grun}, and the first independent solution,
\begin{equation}
p=sCV^{-(s+1)}=C^{\prime}T^{1+1/s}, \qquad \left(\frac{\partial E}{\partial V}\right)_T>0,\label{eq:attractive}
\end{equation}
and the system will condense and form a phase equilibrium. 
\end{enumerate}It is apparent that the bag model is a repulsive equilibrium \eqref{eq:repulsive}, but with an incorrect form of the pressure. 

With the aid of the equation of state, \eqref{eq:Grun}, we can write the first law, \eqref{eq:dE}, in the form
\begin{equation}
dQ=dE+sE\frac{dV}{V}, \label{eq:dE-bis}
\end{equation}
which shows that $V^s$ is an integrating factor, i.e., 
\begin{equation}
V^sdQ=d\left(EV^s\right). \label{eq:int-f}
\end{equation}
Moreover, if we know $dQ=TdS$, then \eqref{eq:int-f} becomes
\begin{equation}
\left(TV^s\right)dS=d\left(EV^s\right). \label{eq:int-f-bis}
\end{equation}
Availing ourselves of \eqref{eq:phi} shows that
\begin{equation}
S=\varphi\left(TV^s\right), \label{eq:S}
\end{equation}
where $\varphi$ is another arbitrary function. The condition of an adiabatic is
\begin{equation}
TV^s=\mbox{const.}, \label{eq:adiabat}
\end{equation}
but if the volume were fixed because the pressure is constant [cf. \eqref{eq:Rmin}] then \eqref{eq:adiabat} would also be an isotherm. 

This would lead to a coincidence of adiabats and isotherms which only happens at absolute zero. Since this is impossible, the volume cannot be fixed by varying the mass in the bag model with respect to its radius. Fixing the volume would lead to fixing the mass. This is the macroscopic analog of there not being a fundamental length for, otherwise, there would be a fundamental mass. This invalidates the bag model~\cite{L07}. 

Moreover, the difference between proton mass and 3 times the quark mass m, in \eqref{eq:M-non}, is identified as the enthalpy, $H/c^2$. The stationary condition, which incorrectly is used to determine the minimum radius, \eqref{eq:Rmin}, is actually the equation of state, \eqref{eq:Grun}, 
\begin{equation}
2\pi r^3=\twothirds pV=\frac{\hbar^2}{2mr^2}=E, \label{eq:eos-bis}
\end{equation}
which determines the pressure as a function of the bag radius. This gives the non-relativistic result that the enthalpy,
\[H=\fivethirds CV^{-\twothirds},\]
is $\fivethirds$ the kinetic energy, \eqref{eq:E-non}.
The volume must be fixed by some non-thermodynamic argument, employing a model, which will then determine the mass.
\section{Bag and Hyperbolic Regge Trajectories}
The bag idea is to consider both mesons and baryons as quarks and antiquarks that are connected by color flux lines. The color fluxes, created by color electrodynamic fields,  terminate at opposite ends of a tube. They supposedly provide a \emph{constant\/} pressure which prevents the walls of the tube from collapsing. However, it is rather hard to imagine how \lq\lq the motions which possess a lot of angular momentum should be long one-dimensional configurations of the bag\rq\rq~\cite{JT}, since both angular momentum and the color magnetic field need at least $2$ dimensions.

The length of the tube $2r_0$, which determines the minimum frequency according to $r_0=c/\omega$, is derived by minimizing the total mass with respect to the length of the tube \lq\lq\emph{at constant total angular momentum\/}\rq\rq~\cite{JT}. The total angular momentum is assumed to be the sum of the angular momentum of the quarks at the ends of the tube, $mcr_0$, and the Poynting vector flux of the color fields, which is quadratic in $r_0$~\cite[Eq. (14)]{JT}. It is hardly credible, therefore, how  the total mass can be made stationary with respect to the length, $2r_0$, at \lq\lq fixed total angular momentum\rq\rq~\cite[Eq. (18)]{JT}.

Moreover, the bag derivation of the Regge trajectories makes use of the volume~\cite{JT}
\begin{align}
V&=\int d^3r\nonumber\\
&=2\int_0^{r_{0}}dr A^{\prime}\nonumber\\
&=2Ar_0\int^{1}_0d\beta\surd\left(1-\beta^2\right)\label{eq:V}
\end{align}
where a rod of length $2r_0$ is considered to contain all the color lines of force between a $q\bar{q}$ pair in the case of mesons, which are situated at the ends of the bar. The relation,
\begin{equation}
\beta=\frac{v}{c}=\frac{r}{r_0}, \label{eq:beta}
\end{equation}
is used together with the relativistic dependency of the area,
\begin{equation}
A^{\prime}=A\surd\left(1-\beta^2\right). \label{eq:A}
\end{equation}

Relation \eqref{eq:beta} follows from the fact that the velocity $v$ is the angular velocity $r\omega$, so that we are not in an inertial frame of reference. The centrifugal acceleration is $r\omega^2$. Thus, the area relation, \eqref{eq:A}, which is supposedly how the area transforms from one inertial frame to another is wrong. Lastly, there is absolutely no meaning to integrating over values of the radius, from $0$ to $r_0=c/\omega$ to define the \lq total\rq\ angular momentum and energy. Rather, we must turn to hyperbolc geometry to obtain the angular momentum-mass squared relation.

The stereographic line element of  hyperbolic geometry for a uniformly rotating disc is~\cite{L11}:
\begin{equation}
ds=\frac{\surd\left(dr^2+r^2d\varphi^2\right)}{(1-r^2\omega^2/c^2)}.\label{eq:metric}
\end{equation}
Since $\varphi$ is a cyclic coordinate, we know that a first integral exists:
\[\frac{r^2\varphi^{\prime}}{(1-r^2\omega^2/c^2)\surd\left(1+r^2\varphi^{\prime\;2}\right)}=J/mc,\]
which is obtained from the stationary condition of the integral of \eqref{eq:metric} with respect to $\varphi^{\prime}$.
Solving for $\varphi^{\prime}=d\varphi/dr=\dot{\varphi}/\dot{r}=\omega/\dot{r}$, where the dot denotes differentiation with respect to time, it is found that  $r$ will be a periodic function of time if and only if the angular momentum is given by:
\begin{equation} 
J=\frac{m_0r^2\omega}{1-(r\omega/c)^2}=m_0\omega r_0^2\sinh^2(\bar{r}/r_0), \label{eq:J}
\end{equation}
which is $m_0\omega/4\pi$ times the area of a hyperbolic circle of radius $\bar{r}$. A subscript \lq$0$\rq\ has been added to the mass to denote that the mass is the rest mass. Expression \eqref{eq:J} is also obtained in the general theory of relativity~\cite{CM}, but, there the line-element~\cite[Eq. (74)]{CM},
\[
ds^2=dr^2+\frac{r^2d\varphi^2}{1-r^2\omega^2/c^2}, \]is wrong since it does not coincide with \eqref{eq:metric}.  

Since the square of the relativistic mass is:
\begin{equation}
m^2=m_0^2\cosh^2(\bar{r}/r_0)=m_0^2\left\{\sinh^2(\bar{r}/r_0)+1\right\}, \label{eq:m}
\end{equation}
 the square of the hyberbolic sine between \eqref{eq:J} and \eqref{eq:m} can be eliminated thus expressing the angular momentum in terms of the square of the mass. We thus obtain the Regge trajectories,
\begin{equation}
J=\alpha^{\prime}m^2-\alpha(0), \label{eq:Regge}
\end{equation}
with the slope,  
\begin{align}
\alpha^{\prime}&=\frac{\omega r_0^2}{m_0}, \label{eq:slope}
\intertext{and intercept,}
\alpha(0)&=J_0=m_0r_0^2\omega. \label{eq:intercept}
\end{align}

The intercept is none other than the Euclidean definition of the angular momentum, and Regge trajectories, \eqref{eq:Regge}, are simply a statement of the conservation of angular momentum in hyperbolic space:
\begin{equation}
\alpha^{\prime}m^2-J=\mbox{const}.\label{eq:J-con}
\end{equation}
The hyperbolic angular momentum, \eqref{eq:J}, vanishes when the relativistic mass coincides with the rest mass, or, equivalently, when $\bar{r}=r_0\tanh^{-1}r/r_0=0$. Expression \eqref{eq:J-con} takes its place next to the relativistic conservation of energy:
\begin{align}
\left(m_0c^2\right)^2&=E^2-(pc)^2\nonumber\\
&=\frac{(mc^2)^2}{1-v^2/c^2}-\frac{(mvc)^2}{1-v^2/c^2}\nonumber\\
&=\left(m_0c^2\right)^2\left\{\cosh^2(\bar{v}/c)-\sinh^2(\bar{v}/c)\right\}, \label{eq:E-con}
\end{align}
in velocity space, with  linear momentum,
\begin{equation}
p=\frac{mv}{\surd(1-v^2/c^2)}=mc\sinh(\bar{v}/c), \label{eq:p}
\end{equation}
and velocity $\bar{v}=c\tanh^{-1}(v/c)$. The connection between the two hyperbolic measures is $\bar{v}=\bar{r}\omega$.

A negative intercept occurs for the $K$ meson with mass $497.7\;(\pm1)$ MeV. The rest mass may be calculated as the square root of the intercept to slope values,
\begin{equation}
\surd\left(\frac{\alpha(0)}{\alpha^{\prime}}\right)\simeq\surd\left(\frac{0.2}{0.8}\right)=500\;\;\mbox{MeV}, \label{eq:ratio}
\end{equation}
which is remarkably close to $497.7(\pm1)$ MeV. Assuming its radius is about $1$ fm, the experimental value for the slope of the $K$-trajectory is $\alpha^{\prime}=0.8$ GeV$^{-2}$, so that \eqref{eq:slope} gives a frequency, $\omega=1.24\times10^{22}$ Hz. This is some $10^8$ larger than the frequencies of the velocity of visible light. The intercept can then be determined to be $\alpha(0)=0.225$, which is in agreement with the experimental value of $\alpha(0)=0.2$~\cite{PDBC}. The angular momentum \eqref{eq:Regge} vanishes, thus, placing $K$'s squared mass on the axis of the plot, $J$ versus $m^2$.

The bag derivation of the Regge trajectories  determines the frequency of the rotating meson by determining $r_0$ through a mimimization of the total mass at constant angular momentum~\cite{JT}. However, the \lq total\rq\ angular momentum should also be a function of $r_0$ because it is the moment of the momentum. The supposed integration of the Poynting vector has camouflaged that there is an integration of all frequencies up to $c/r_0$, since the frequency appears together with $r$ in the form $\beta=r\omega/c$. So the integral over $r$ from $0$ to $r_0$ can equally as well be interpreted as an integral over $\omega$ from $0$ to $c/r_0$.  Integrating the angular momentum over $r$ or $\omega$ is meaningless since it gives a quadratic dependence of $r_0$ on the angular momentum. This means that the angular momentum is inversely proportional to $\omega^2$ leading to the absurd conclusion that the angular momentum decreases with increasing frequency. 

This renders invalid the derivation of the Regge trajectories in ref.~\cite{bhl}, as well as all the references from which it was taken. Finally, the bag model cannot determine the all important intercept, which is the discriminating factor of the different families of rotational excitations, since their slopes are all in the small interval from $0.8$ to $0.9$ GeV$^{-2}$.

\end{document}